\documentclass{llncs}

\usepackage{url}

\usepackage{verbatim}
\usepackage{amssymb}
\usepackage{rotating}
\usepackage{booktabs}

\usepackage{array}

\usepackage{color}

\title{Finding Quality Issues in SKOS Vocabularies}
\author{Christian Mader\inst{1} \and Bernhard Haslhofer\inst{2} \and Antoine Isaac\inst{3}}
\institute{
	University of Vienna, Faculty of Computer Science, Austria\\\email{christian.mader@univie.ac.at}
	\and Cornell University, Department of Information Science, USA\\\email{bernhard.haslhofer@cornell.edu}
	\and Europeana \& Vrije Universiteit Amsterdam, The Netherlands\\\email{aisaac@few.vu.nl}
	}

\begin{document}

\maketitle

\begin{abstract}
    
The Simple Knowledge Organization System (SKOS) is a standard model for controlled vocabularies on the Web. However, SKOS vocabularies often differ in terms of quality, which reduces their applicability across system boundaries. Here we investigate how we can support taxonomists in improving SKOS vocabularies by pointing out quality issues that go beyond the integrity constraints defined in the SKOS specification. We identified potential quantifiable quality issues and formalized them into computable quality checking functions that can find affected resources in a given SKOS vocabulary. We implemented these functions in the qSKOS quality assessment tool, analyzed 15 existing vocabularies, and found possible quality issues in all of them.

\end{abstract}

\setcounter{footnote}{0}


\section{Introduction}\label{sec:introduction}


The Simple Knowledge Organization System (SKOS)~\cite{SkosReference2008} is a standard model for sharing and linking controlled vocabularies (thesauri, classification systems, etc.) on the Web. Organizations like, e.g., the European Union\footnote{EuroVoc, \url{http://eurovoc.europa.eu/}}, the United Nations\footnote{AGROVOC, \url{http://aims.fao.org/standards/agrovoc/about}}, or the UK government\footnote{Integrated Public Sector Vocabulary (IPSV), \url{http://doc.esd.org.uk/IPSV}} publish SKOS representations of their vocabularies on the Web so that they can easily be accessed by humans and machines.

However, quality issues can affect the applicability of SKOS vocabularies for tasks such as query expansion, faceted browsing, or auto-completion, as in the following examples: 

\begin{itemize}

	\item AGROVOC defines concepts in 25 different languages. While most concepts have English labels attached, only 38\% have German labels. This can be a problem for multilingual applications that rely on label translations.

	\item An earlier version of the STW thesaurus (v8.06) contained 5 pairs of concepts with identical labels. As a result, the auto-complete function of the online search interface suggested identical entries without disambiguation information.

	\item The non-public thesaurus of the Austrian Armed Forces (LVAk) contains 11 disconnected concept clusters. When confronted with these structures, the maintainers recognized them as data without practical significance.
	
\end{itemize}


The SKOS specification defines a set of integrity conditions that state whether given data patterns are consistent with the SKOS model. Yet the SKOS integrity conditions fail to capture quality aspects like the ones above. The main reason lies in SKOS' ``minimal commitment'' approach. A standard that aims at cross-domain interoperability should refrain from defining constraints that impose on one domain the requirements of another. SKOS is thus very liberal with respect to data quality. On the other hand, each vocabulary should fulfill domain- and application-specific quality aspects and taxonomists often follow standard guidelines specific to given types of vocabularies (cf.,~\cite{ISO25964-1:2011,Z39.19:2005}) or apply their own hand-crafted checks~\cite{Coronado2009}. Existing guidelines consider these aspects, but currently rely on human judgment, which is subjective and does not scale for larger vocabularies. The SKOS context, where vocabularies can be linked together on the Web, also brings issues hitherto unforeseen by traditional checking approaches.


We aim at contributing to the ongoing community efforts to bridge that gap between model-level integrity constraints and domain-specific quality aspects. Our goal is to help taxonomists in identifying possible quality issues in SKOS vocabularies and to give them a set of computable quality checking functions that, in combination with the taxonomists' experience and domain expertise, can serve as quality indicators for vocabularies. Finding such quality issues also gives important feedback on the overall vocabulary design process and should, in the end, lead to better vocabularies. With defining computable quality checking functions we tackle the problem of quality assessment from an objective perspective. Subjective perception of quality is not within the scope of this work. Our contribution can be summarized as follows:

\begin{itemize}

	\item We identified 15 quality issues for SKOS vocabularies by examining existing guidelines and formalized them into computable quality checking functions that identify possibly affected resources in a vocabulary.
	
	\item With the qSKOS quality assessment tool we provide a reference implementation of these functions.

	\item We tested these functions by analyzing a representative set of 15 existing SKOS vocabularies to learn about possible quality issues.

\end{itemize}

In the following, we will first discuss what ``quality'' means in the context of SKOS vocabularies and how it is currently supported by the SKOS specification and existing tools. Then we introduce the quality issues we have identified and describe how we implemented them in the qSKOS quality assessment tool. Finally, we report on the results of an analysis we performed on 15 existing SKOS vocabularies and show that the quality issues we discussed are real and can lead to the improvement of existing vocabularies. All supplemental materials\footnote{qSKOS: \url{https://github.com/cmader/qSKOS/}, wiki: \url{https://github.com/cmader/qSKOS/wiki}, dataset: \url{https://github.com/cmader/qSKOS-data/}} are available online.


\section{Background and Related Work}\label{sec:background}

The problem of ``vocabulary quality'' is closely related to the more general one of ``data quality'' and has been discussed in data and information systems research (cf.~\cite{Batini2009}). Pipino et al.~\cite{Pipino2002} argue that dealing with data quality should involve both ``subjective perceptions of the individuals'' and ``objective measurements based on the data''. We see our work as a contribution to the latter.

The SKOS specification does not mention the notion of quality, but defines in total six integrity conditions~\cite{SkosReference2008}, each of which is a statement that defines under which circumstances data are consistent with the SKOS data model. For example, ``a resource has no more than one value of \texttt{skos:prefLabel} per language tag''. Tools that can check whether these conditions are met are already available: two of the six conditions are defined formally in the OWL representation of SKOS and can therefore be validated by any OWL reasoner. For validating a SKOS vocabulary against the other integrity conditions, one can use tools such as the PoolParty Thesaurus Consistency Checker\footnote{\url{http://demo.semantic-web.at:8080/SkosServices/check}}, or the Skosify\footnote{\url{http://code.google.com/p/skosify/}} validator, which can also correct some detected quality problems.

Typical applications of controlled vocabularies are classification, indexing, auto-completion, query reformulation, or serving as a glossary. As we discussed in detail in earlier work~\cite{Nagy2011}, these areas impose specific requirements on vocabulary features, such as structure, availability, and documentation. Quality aspects of controlled vocabularies have already been discussed in standardized guidelines~\cite{ISO25964-1:2011,Z39.19:2005}, manuals~\cite{Aitchison2000,Harpring2010,Hedden2010,Svenonius2003}, tutorials~\cite{Soergel2002}, and scholarly articles \cite{Coronado2009,Kless2010}. These most often rely on manual, precise analysis of individual statements in the data, as in~\cite{spero2008}. Our work builds on this literature, but focuses on the less intellectually loaded checks, which can be automatized to assist vocabulary users or publishers. 


Data quality is also being discussed in Semantic Web and Linked Data research. Hogan et. al~\cite{Hogan2010} identify four categories of common errors and shortcomings in RDF documents and Heath and Bizer~\cite{Heath2011} summarize best practices for publishing data on the Web. Ontology evaluation, i.e., measuring the quality of an ontology, has also been discussed extensively~\cite{Vrandecic2010}. However, the authors focus on RDF datasets and ontologies in general. While we could use some criteria suggested here, such as consistent tagging of literals, these need to be completed by considering SKOS-specific properties.

One issue when assessing the quality of SKOS data is the so-called ``Open World Assumption'', which underlies the Web of Data itself. Established quality notions from closed-world systems, such as referential integrity or schema validation, do not hold anymore, because available information may be incomplete and non-explicitly stated facts cannot be determined as true or false. Work-arounds, often ad-hoc, are thus currently used to evaluate quality in Linked Data sets~\cite{Heath2011}, as is done in the (rule-based) SKOS tools mentioned above, or the Pellet ICV\footnote{\url{http://clarkparsia.com/pellet/icv/}}, which re-interprets OWL axioms with integrity constraint semantics.


\section{Quality Issues in SKOS Vocabularies}\label{sec:criteria}


We identified an initial set of possible quality issues in SKOS vocabularies by reviewing literature, manually examining existing vocabularies, and focusing on issues that can be measured automatically. Some measures, such as hierarchy depth or node centrality, have been omitted due to lack of evidence on their general influence on vocabulary quality.
We published our findings in the qSKOS wiki and requested feedback from experts via public mailing lists and informal face to face discussions. Based on the received responses, we translated a subset of these issues into computable quality checking functions. Each function takes a given SKOS vocabulary and an optional vocabulary namespace as input and finds all resources that match the corresponding quality issue. For the purpose of this work, we define a SKOS vocabulary as follows:

\spnewtheorem*{mydef}{Definition}{\bfseries}{\itshape}
\begin{mydef}[SKOS Vocabulary]
Let a SKOS vocabulary be a tuple of the form $V = \langle IR, C, AC, SR, LV, CS \rangle$, with \(IR = I_{CEXT}(\texttt{rdfs:Resource}^\mathcal{I})\) being the set of \textbf{resources}, \(C \subseteq IR\) with \(C = I_{CEXT}(\texttt{skos:Concept}^\mathcal{I})\) being the set of \textbf{concepts}, \(AC \subseteq C\) being the set of \textbf{authoritative concepts}, which are all concepts that are identified by URIs in the vocabulary namespace, \(SR = I_{EXT}(\texttt{skos:semanticRelation}^\mathcal{I})\) being the set of \textbf{semantic relations} associating concepts with one another, $LV \subseteq I_{CEXT}(\texttt{rdfs:Literal}^\mathcal{I})$ being the set of untyped \textbf{plain literals}, and \(CS = I_{CEXT}(\texttt{skos:ConceptScheme}^\mathcal{I})\) being the set of \textbf{concept schemes}. Further, we let $V$ be the fully entailed RDFS interpretation of the underlying RDF graph. We enrich $V$ by entailment of \texttt{owl:inverseOf} properties as well as instances of \texttt{owl:TransitiveProperty} and \texttt{owl:SymmetricProperty} defined by the formal OWL semantics of SKOS~\cite{SkosReference2008}.
\end{mydef}

In the following, we explain the origins and design rationale for each quality issue and explain how the corresponding quality checking function works. For better readability and due to lack of space we provide only semi-formal definitions and refer to the source code of the qSKOS tool for further details.



\subsection{Labeling and Documentation Issues}

\paragraph{Omitted or Invalid Language Tags}

SKOS defines a set of properties that link resources with RDF Literals, which are plain text natural language strings with an optional language tag. This includes the labeling properties {\texttt{rdfs:label}, \texttt{skos:prefLabel}, \texttt{skos:altLabel}, \texttt{skos:hiddenLabel} and also SKOS documentation properties, such as \texttt{skos:note} and subproperties thereof. Literals should be tagged consistently~\cite{Vrandecic2010}, because omitting language tags or using non-standardized, private language tags in a SKOS vocabulary could unintentionally limit the result set of language-dependent queries.
A SKOS vocabulary can be checked for omitted and invalid language tags by iterating over all resources in $IR$ and finding those that have labeling or documentation property relations to plain literals in $LV$ with missing or invalid language tags, i.e., tags that are not defined in RFC3066\footnote{Tags for the Identification of Languages \url{http://www.ietf.org/rfc/rfc3066.txt}}.

\paragraph{Incomplete Language Coverage}

The set of language tags used by the literal values linked with a concept should be the same for all concepts. If this is not the case, appropriate actions like, e.g., splitting concepts or introducing scope notes should be taken by the creators. This is particularly important for applications that rely on internationalization and translation use cases.
Affected concepts can be identified by first extracting the global set of language tags used in a vocabulary from all literal values in $LV$, which are attached to a concept in $C$. In a second iteration over all concepts, those having a set of language tags that is not equal to the global language tag set are returned.

\paragraph{Undocumented Concepts}

Svenonius~\cite{Svenonius1997} advocates the ``inclusion of as much definition material as possible'' and the SKOS Reference~\cite{SkosReference2008} defines a set of ``documentation properties'' intended to hold this kind of information.
To identify all undocumented concepts, we iterate over all concepts in $C$ and collect those that do not use any of these documentation properties.

\paragraph{Label Conflicts}

The SKOS Primer~\cite{Isaac2009} recommends that ``no two concepts have the same preferred lexical label in a given language when they belong to the same concept scheme''. This issue could affect application scenarios such as auto-completion, which proposes labels based on user input. Although these extra cases are acceptable for some thesauri, we generalize the above recommendation and search for all concept pairs with their respective \texttt{skos:prefLabel}, \texttt{skos:altLabel} or \texttt{skos:hiddenLabel} property values meeting a certain similarity threshold defined by a function $sim:LV \times LV \rightarrow [0,1]$.
The default, built-in similarity function checks for case-insensitive string equality with a threshold equal to 1. Label conflicts can be found by iterating over all (authoritative) concept pairs $AC \times AC$, applying $sim$ to every possible label combination, and collecting those pairs with at least one label combination meeting or exceeding a specified similarity threshold. We handle this issue under the Closed World Assumption, because data on concept scheme membership may lack and concepts may be linked to concepts with similar labels in other vocabularies.


\subsection{Structural Issues}

\paragraph{Orphan Concepts}

are motivated by the notion of ``orphan terms'' in the literature~\cite{Hedden2010}, i.e., terms without any associative or hierarchical relationships. Checking for such terms is common in thesaurus development and also suggested by~\cite{Z39.19:2005}. Since SKOS is concept-centric, we understand an orphan concept as being a concept that has no semantic relation $sr \in SR$ with any other concept. Although it might have attached lexical labels, it lacks valuable context information, which can be essential for retrieval tasks such as search query expansion.
Orphan concepts in a SKOS vocabulary can be found by iterating over all elements in $C$ and selecting those without any semantic relation to another concept in $C$.

\paragraph{Weakly Connected Components}

A vocabulary can be split into separate ``clusters'' because of incomplete data acquisition, deprecated terms, accidental deletion of relations, etc. This can affect operations that rely on navigating a connected vocabulary structure, such as query expansion or suggestion of related terms.
Weakly connected components are identified by first creating an undirected graph that includes all non-orphan concepts (as defined above) as nodes and all semantic relations $SR$ as edges. ``Tarjan's algorithm''~\cite{Hopcroft1973} can then be applied to find all connected components, i.e., all sets of concepts that are connected together by (chains of) semantic relations.

\paragraph{Cyclic Hierarchical Relations}

is motivated by Soergel et al.~\cite{Soergel2002} who suggest a ``check for hierarchy cycles'' since they ``throw the program for a loop in the generation of a complete hierarchical structure''. Also Hedden~\cite{Hedden2010}, Harpring~\cite{Harpring2010} and Aitchison et al.~\cite{Aitchison2000} argue that there exist common forms like, e.g., ``generic-specific'', ``instance-of'' or ``whole-part'' where cycles would be considered a logical contradiction.
Cyclic relations can be found by constructing a graph with the set of nodes being $C$ and the set of edges being all \texttt{skos:broader} relations.

\paragraph{Valueless Associative Relations}

The ISO/DIS 25964-1 standard \cite{ISO25964-1:2011} suggests that terms that share a common broader term should not be related associatively if this relation is only justified by the fact that they are siblings. This is advocated by Hedden~\cite{Hedden2010} and Aitchison et al.~\cite{Aitchison2000} who point out ``the risk that thesaurus compilers may overload the thesaurus with valueless relationships'', having a negative effect on precision.
This issue can be checked by identifying concept pairs $C \times C$ that share the same broader or narrower concept while also being associatively related by the property \texttt{skos:related}.

\paragraph{Solely Transitively Related Concepts}

Two concepts that are explicitly related by \texttt{skos:broaderTransitive} and/or \texttt{skos:narrowerTransitive} can be regarded a quality issue because, according to \cite{SkosReference2008}, these properties are ``not used to make assertions''. Transitive hierarchical relations in SKOS are meant to be inferred by the vocabulary consumer, which is reflected in the SKOS ontology by, for instance, \texttt{skos:broader} being a subproperty of \texttt{skos:broaderTransitive}. 
This issue can be detected by finding all concept pairs $C \times C$ that are directly related by \texttt{skos:broaderTransitive} and/or \texttt{skos:narrowerTransitive} properties but not by (chains of) \texttt{skos:broader} and \texttt{skos:narrower} subproperties.

\paragraph{Omitted Top Concepts}

The SKOS model provides concept schemes, which are a facility for grouping related concepts. This helps to provide ``efficient access''~\cite{Isaac2009} and simplifies orientation in the vocabulary. In order to provide entry points to such a group of concepts, one or more concepts can be marked as top concepts.  
Omitted top concepts can be detected by iterating over all concept schemes in $CS$ and collecting those that do not occur in relations established by the properties \texttt{skos:hasTopConcept} or \texttt{skos:topConceptOf}.

\paragraph{Top Concept Having Broader Concepts}

Allemang et al. \cite{Allemang2011} propose to ``not indicate any concepts internal to the tree as top concepts'', which means that top concepts should not have broader concepts. 
Affected resources are found by collecting all top concepts that are related to a resource via a \texttt{skos:broader} statement and not via \texttt{skos:broadMatch}---mappings are not part of a vocabulary's ``intrinsic'' definition and a top concept in one vocabulary may perfectly have a broader concept in another vocabulary.


\subsection{Linked Data Specific Issues}\label{subsec:ld_issues}

\paragraph{Missing In-Links}

When vocabularies are published on the Web, SKOS concepts become linkable resources. Estimating the number of in-links and identifying the concepts without any in-links, can indicate the importance of a concept.
We estimate the number of in-links by iterating over all elements in $AC$ and querying the Sindice\footnote{\url{http://sindice.com/} indexes the Web of Data, which is composed of pages with semantic markup in RDF, RDFa, Microformats, or Microdata. Currently it covers approximately 230M documents with over 11 billion triples.} SPARQL endpoint for triples containing the concept's URI in the object part. Empty query results are indicators for missing in-links.

\paragraph{Missing Out-Links}

SKOS concepts should also be linked with other related concepts on the Web, ``enabling seamless connections between data sets''\cite{Heath2011}. Similar to \emph{Missing In-Links}, this issue identifies the set of all authoritative concepts that have no out-links.
It can be computed by iterating over all elements in $AC$ and returning those that are not linked with any non-authoritative resource.

\paragraph{Broken Links}

As we discussed in detail in our earlier work~\cite{Popitsch:2010:DHB:1772690.1772768}, this issue is caused by vocabulary resources that return HTTP error responses or no response when being dereferenced. An erroneous HTTP response in that case can be defined as a response code other than 200 after possible redirections. Just as in the ``document'' Web, these ``broken links'' hinder navigability also in the Linked Data Web and and should therefore be avoided. 
Broken links are detected by iterating over all resources in $IR$, dereferencing their HTTP URIs, following possible redirects, and including unavailable resources in the result set.

\paragraph{Undefined SKOS Resources}

The SKOS model is defined within the namespace \url{http://www.w3.org/2004/02/skos/core#}. However, some vocabularies use resources from within this namespace, which are unresolvable for two main reasons: vocabulary creators ``invented'' new terms within the SKOS namespace instead of introducing them in a separate namespace, or they use ``deprecated'' SKOS elements like \texttt{skos:subject}.
Undefined SKOS resources can be identified by iterating over all resources in $IR$ and returning those (i) that are contained in the list of deprecated resources\footnote{See \url{http://www.w3.org/TR/skos-reference/#namespace}} or (ii) are identified by a URI in the SKOS namespace but are not defined in the current version of the SKOS ontology.


\section{Analysis of Existing SKOS Vocabularies}\label{sec:analysis}

We used the qSKOS quality assessment tool to find possible quality issues in existing SKOS vocabularies. From each quality checking function we obtained detailed reports listing possibly affected resources.

\subsection{Vocabulary Data Set}

Table~\ref{tab:vocabs} summarizes some basic statistical properties of our vocabulary selection: the number of concepts and authoritative concepts, all \texttt{skos:prefLabel}, \texttt{skos:altLabel}, and \texttt{skos:hiddenLabel} relations involving concepts (\texttt{Concept Labels}), all asserted semantic relations, as well as the number of concept schemes. From these properties we can, for instance, already see that approximately 3,000 \texttt{DBpedia Categories} concepts do not have labels (e.g., \url{Category:South_Korean_social_scientists}), which is an indicator for missing natural language descriptions in some Wikipedia categories. 

\begin{table}
\caption{Analyzed SKOS vocabularies}
\label{tab:vocabs}
\begin{center}
\resizebox{12cm}{!} {
\setlength{\extrarowheight}{5pt}

\begin{tabular}{p{6cm}ccccccccc}

\textbf{Vocabulary} & \rotatebox{90}{\textbf{Abbreviation}} & \rotatebox{90}{\textbf{\parbox{2.5cm}{Version/\\Last Modified}}} & \rotatebox{90}{\textbf{Concepts}} & \rotatebox{90}{\textbf{\parbox{2.5cm}{Authoritative\\Concepts}}} & \rotatebox{90}{\textbf{\parbox{2.5cm}{Concept\\Labels}}} & \rotatebox{90}{\textbf{\parbox{2.5cm}{Semantic\\Relations}}} & \rotatebox{90}{\textbf{\parbox{2.53cm}{Concept\\Schemes}}} \\
\toprule
United Nations Agricultural Thesaurus & \textbf{AGROVOC} & 1.3 & 32,035 & 32,035 & 620,629 & 65,934 & 1 \\
\hline
DBpedia Categories & \textbf{DBpedia} & 3.7 & 743,410 & 743,410 & 740,352 & 1,490,316 & 0 \\
\hline
The EU's Multilingual Thesaurus & \textbf{Eurovoc} & 5.0 & 6,797 & 6,797 & 457,788 & 18,491 & 128 \\
\hline
Geonames Ontology & \textbf{Geonames} & 2.2.1 & 671 & 671 & 671 & 0 & 9 \\
\hline
Gemeenschappelijke Thesaurus Audiovisuele Archieven & \textbf{GTAA} & 2010/08/25 & 171,991 & 171,991 & 178,776 & 50,892 & 9 \\
\hline
Integrated Public Sector Vocabulary & \textbf{IPSV} & 2.00 & 4,732 & 3,080 & 7,945 & 13,843 & 3 \\
\hline
Library of Congress Subject Headings & \textbf{LCSH} & 2012/03/29 & 443,164 & 408,009 & 750,219 & 598,134 & 1 \\
\hline
Austrian Armed Forces Thesaurus & \textbf{LVAk} & 0.9 & 13,411 & 13,411 & 17,250 & 16,346 & 0 \\
\hline
Middle Kingdom Tombs of Ancient Egypt Thesaurus & \textbf{Meketre} & 2011/07/07 & 422 & 422 & 569 & 1,698 & 2 \\
\hline
Medical Subject Headings & \textbf{MeSH} & \cite{van2006method} & 24,626 & 24,626 & 150,617 & 38,858 & 0 \\
\hline
North American Industry Classification System & \textbf{NAICS} & 2012 & 4,175 & 2,213 & 0 & 8,684 & 1 \\
\hline
New York Times People & \textbf{NYTP} & 2010/06/22 & 4,979 & 4,979 & 4,979 & 0 & 1 \\
\hline
University of Southampton Pressinfo & \textbf{Pressinfo} & 2011/02/24 & 1,125 & 1,125 & 0 & 0 & 0 \\
\hline
Peroxisome Knowledge Base & \textbf{PXV} & 1.6 & 2,112 & 1,686 & 3,628 & 2,695 & 1 \\
\hline
Thesaurus for Economics & \textbf{STW} & 8.10 & 25,107 & 6,789 & 58,441 & 91,816 & 3 \\
\bottomrule
\end{tabular}
}
\end{center}
\end{table}

\subsection{Results and Discussion}

The results of this analysis are summarized in Table~\ref{tab:results}, which shows the absolute number of possibly affected resources for each quality checking function and vocabulary. Numbers marked with an asterisk (*) were obtained by extrapolating from subsets containing 5\% of the respective vocabulary resources.

\vspace{-.3cm}
\begin{table}[h]
\caption{Results of the quality checking functions}
\label{tab:results}
\vspace{-.5cm}
\begin{center}
\resizebox{\textwidth}{!} {
\setlength{\extrarowheight}{5pt}

\begin{tabular}{lccccccccccccccc}
\textbf{Issue} & \rotatebox{90}{\textbf{AGROVOC}} & \rotatebox{90}{\textbf{DBpedia}} & \rotatebox{90}{\textbf{Eurovoc}} & \rotatebox{90}{\textbf{Geonames}} & \rotatebox{90}{\textbf{GTAA}} & \rotatebox{90}{\textbf{IPSV}} & \rotatebox{90}{\textbf{LCSH}} & \rotatebox{90}{\textbf{LVAk}} & \rotatebox{90}{\textbf{Meketre}} & \rotatebox{90}{\textbf{MeSH}} & \rotatebox{90}{\textbf{NAICS}} & \rotatebox{90}{\textbf{NYTP}} & \rotatebox{90}{\textbf{Pressinfo}} & \rotatebox{90}{\textbf{PXV}} & \rotatebox{90}{\textbf{STW}} \\

\toprule
Omitted or Invalid Language Tags & 0 & 0 & 219 & 0 & 0 & 0 & 100,316 & 13,411 & 0 & 23,950 & 0 & 0 & 1,224 & 1,578 & 2 \\

Incomplete Language Coverage & 32,035 & 0 & 6370 & 0 & 0 & 0 & 0 & 0 & 420 & 0 & 0 & 0 & 0 & 0 & 25,050 \\

Undocumented Concepts & 32,035 & 743,410 & 5,341 & 0 & 96,850 & 4,551 & 342,848 & 13,411 & 422 & 1,807 & 3,259 & 4,094 & 1,125 & 1,918 & 23,752 \\

Label Conflicts & 2,949 & 0 & 48 & 18 & 12,404 & 0 & 10,862 & 13 & 4 & 0 & 0 & 0 & 0 & 7 & 0 \\

\midrule

Orphan Concepts & 0 & 77,062 & 7 & 671 & 162,000 & 0 & 173,149 & 21 & 0 & 0 & 0 & 4,979 & 1,125 & 2 & 70 \\

Weakly Connected Components & 4 & 1,506 & 4 & 0 & 621 & 1 & 22,343 & 11 & 5 & 4 & 1 & 0 & 0 & 10 & 141 \\

Cyclic Hierarchical Relations & 0 & 1,132 & 0 & 0 & 0 & 0 & 0 & 5 & 0 & 4 & 0 & 0 & 0 & 0 & 0 \\

Valueless Associative Relations & 282 & 8,839 & 6 & 0 & 9,448 & 253 & 0 & 5 & 0 & 550 & 0 & 0 & 0 & 0 & 5,004 \\

Solely Transitively Related Concepts & 0 & 0 & 2,652 & 0 & 0 & 0 & 0 & 0 & 36 & 0 & 2,189 & 0 & 0 & 0 & 0 \\

Omitted Top Concepts & 0 & 0 & 1 & 9 & 9 & 0 & 1 & 0 & 0 & 0 & 0 & 1 & 0 & 0 & 2 \\

Top Concept Having Broader Concepts & 0 & 0 & 0 & 0 & 0 & 0 & 0 & 0 & 0 & 0 & 0 & 0 & 0 & 1 & 0 \\

\midrule

Missing In-Links & 32,035 & 733,800 & 6,796 & 19 & 171,980* & 3,080 & 408,000* & 13,411 & 422 & 24,625 & 2,213 & 20 & 1,125 & 1,686 & 6,781 \\

Missing Out-Links & 32,035 & 743,410 & 6,797 & 671 & 171,991 & 0 & 344,054 & 13,411 & 273 & 24,626 & 1 & 0 & 1,116 & 1,046 & 0 \\

Broken Links & 238 & 0* & 0* & 0 & 0 & 1 & 780 & 0 & 425 & 1 & 3,169 & 7 & 11 & 163 & 575 \\

Undefined SKOS Resources & 0 & 0 & 0 & 0 & 0 & 1 & 0 & 0 & 0 & 1 & 0 & 0 & 0 & 0 & 0 \\

\bottomrule
\end{tabular}
}
\end{center}
\end{table}
\vspace{-.3cm}

We found labeling and documentation issues in all vocabularies.
\texttt{MeSH}, \texttt{PXV}, \texttt{Pressinfo}, and \texttt{LVAk} omit language tags with their labeling properties, \texttt{LCSH} with the \texttt{skos:note} property. \texttt{STW} does not use language tags with 2 instances of \texttt{skos:definition}.
\texttt{AGROVOC} covers 25 languages but no single concept is labeled in all languages, in \texttt{Meketre} all concepts have English but only some of them French labels assigned. \texttt{STW}, which is expressed mainly in English and German, has many concepts with incomplete language coverage because it (i) links to non-authoritative concepts that are only labeled in German and (ii) uses the private, but valid language tag \texttt{x-other} with some of its concept labels.
\texttt{Geonames}, which defines a concept scheme of ``feature codes'', is the only vocabulary in our dataset, which has at least one documentation property assigned to all of its concepts. All other vocabularies have a significant number of undocumented concepts.
We also detected possible label conflicts in half of the vocabularies. \texttt{PXV}, for instance, uses the string ``primary peroxisomal enzyme deficiency'' with two concepts in the same concept scheme, but once with a \texttt{skos:prefLabel} and another time with a \texttt{skos:altLabel} property. In \texttt{NAICS} we could not detect any labeling issues but found that it expresses statements with \texttt{skosxl:prefLabel} as predicate and plain literals as object, which contradicts the SKOS-XL\footnote{The SKOS eXtension for Labels (SKOS-XL) provides additional support for identifying, describing and linking lexical entities.~\cite{SkosReference2008}} specification. 

When analyzing the vocabularies for structural issues, we found that certain results can be seen as indicators for the types of vocabularies.
In the \texttt{Pressinfo}, \texttt{Geonames}, and \texttt{NYTP} vocabulary, all concepts are orphan concepts, which means that these vocabularies are authority files rather than thesauri or taxonomies. This also implies that these vocabularies have no weakly connected components. \texttt{GTAA} is a mixture of name authority file (approx.~162K concepts) and thesaurus (approx.~10K concepts). The 70 orphan concepts in \texttt{STW} are deprecated concepts.


Three vocabularies show no weakly connected components (WCCs) because all concepts are orphan concepts and thus no relations between them are established. Two vocabularies (\texttt{IPSV}, \texttt{NAICS}) consist of only one ``giant component'', which is often considered the ideal vocabulary structure. \texttt{STW} forms one giant component (containing 24,572 concepts), but has also 140 additional WCCs, which all contain linked authoritative and non-authoritative concepts. All other vocabularies split into several clusters of semantically related concepts, each of which represents a certain subtopic. \texttt{Eurovoc}, for instance, has 4 WCCs, containing 4, 5, 6 and 6775 concepts. 
In the large WCC it uses a custom ontology to organized numerous micro-thesauri and domains and cross-connects concepts by \texttt{skos:related} properties. However, this is not the case for the three small WCCs, indicating a quality flaw. WCCs divide the \texttt{Meketre} vocabulary into different topics, e.g., museums or concepts reserved for internal use. \texttt{GTAA} consists of 621 highly unbalanced WCCs. One component contains 8413 subjects from a thesaurus with carefully curated semantic relations. Most of the other components contain less than 10 entities from other categories, e.g., locations, person names, and genres, for which the ``traditional'' information management practices involve much less explicit linking. \texttt{PXV} splits into 10 topic-related WCCs, such as ``deficiencies'', ``defects'' or ``signals''. Some of the 11 concept clusters contained in the \texttt{LVAk} thesaurus are obviously forgotten test data. 

Hierarchical cycles are not a common issue except in the collaboratively created \texttt{DBpedia} vocabulary, where many concepts have reflexive \texttt{skos:broader} relations. The cycles in \texttt{MeSH} and \texttt{LVAk} could, in our opinion, be resolved by replacing hierarchical with associative relations or synonym definitions.
Valueless associative relations occur in 8 vocabularies, with their total number being relatively low compared to the total number of all semantic relations in the respective vocabularies.
Solely transitive related concepts occur in 3 vocabularies, establishing relations using properties that, according to~\cite{SkosReference2008}, should not be asserted directly. This indicates a possible misinterpretation of the SKOS specification and could result in a loss in recall on hierarchical queries.
\texttt{GTAA} and \texttt{Geonames} omit top concepts in all concept schemes they define. \texttt{Eurovoc} uses 128 concept schemes but has one without top concept, which simply contains all concepts defined in the vocabulary. Such an ``umbrella concept scheme'' without top concept is also present in \texttt{LCSH} and \texttt{NYTP}.
Only the \texttt{PXV} vocabulary is affected by top concepts having broader concepts in its current version. In earlier versions more of them could be found which were, according to the vocabulary creator, abandoned but still available in the triple store, probably caused by some bug in the vocabulary management software.


The difference between the number of concepts and the number of authoritative concepts in Table~\ref{tab:vocabs} already indicates which vocabularies are linked with other SKOS vocabularies.
However, except \texttt{NYTP} and \texttt{Geonames}, no vocabulary has a high number of estimated in-links from other web resources.
Also the number of out-links is rather low: \texttt{NYTP}, \texttt{IPSV}, and \texttt{STW} are the three exceptions, which are fully linked to other Web resources. The one concept with missing out-links in \texttt{NAICS} is the object of a \texttt{skos:broaderTransitive} relation. One reason for a high number of missing out-links is that the links were not available in the main thesaurus file, which is at least the case for \texttt{AGROVOC}.
Even though we could not determine the exact number of broken links because of the large number of links to resolve  (over 400K in \texttt{Eurovoc}, over 500K in \texttt{LCSH}), we found that broken links are a common issue in most vocabularies.
Undefined SKOS resources seem to be a minor issue, because we could only find two of them in all vocabularies: \texttt{MeSH} introduces \texttt{skos:annotation} and \texttt{IPSV} uses the deprecated \texttt{skos:prefSymbol} property.

\section{Conclusions and Future Work}\label{sec:conclusions}

We presented possible quality issues in SKOS vocabularies and described how we implemented them as quality checking functions in our qSKOS quality assessment tool. We analyzed a representative set of existing SKOS vocabularies and found  issues in all of them. Labeling and documentation issues were omnipresent and also structural issues, which require further investigation by the vocabulary maintainers, were found in most vocabularies. Although SKOS is designed for Linked Data, many existing vocabularies still resemble their closed-system origin, which results in a relatively low number of in- and out-links. Broken links are a major issue and call for synchronization mechanisms in order to maintain navigability between concepts in different vocabularies.

We are aware that these issues are purely quantitative quality indicators. To learn more about the real-world impact of our work like, e.g., the relative importance of the identified quality issues, we will conduct a qualitative follow-up study, in which we discuss these results with more taxonomists. We will also set up a Web-based SKOS quality checking service to further collect community feedback and enhance the issues list. These enhancements may encompass a finer-grained evaluation on some issues like, e.g., documentation quality indicators which could also include average number or length of documentation statements and their standard deviation across concepts.

We already reported initial results from our analysis to some of the maintainers of the vocabularies we analyzed. At the time of this writing, we know that our findings led to improvements in at least two SKOS vocabularies.

\subsection*{Acknowledgements}

We thank Andrew Gibson and Tom Dent for providing PXV and IPSV data and Joachim Neubert for his valuable feedback. The work is supported by the FWF P21571 Meketre project and EU Marie Curie Fellowship no. 252206.




\bibliography{references}
\bibliographystyle{splncs03}

\end{document}